\documentclass{emulateapj}

\slugcomment{Draft, \today}

\def\be{\begin{equation}}
\def\ee{\end{equation}}
\def\nms{\mathsurround=0pt}
\def\oversim#1#2{\lower 4pt\vbox{\baselineskip 0pt \lineskip 1pt
    \ialign{$\nms#1\hfil##\hfil$\crcr#2\crcr\sim\crcr}}}
\def\ga{\mathrel{\mathpalette\oversim>}} 
\def\la{\mathrel{\mathpalette\oversim<}} 
\def\arcdeg{{^{\circ}}}

\def\bh{M_{\bullet}}
\def\bhsp{M_{\bullet,1}}
\def\bhs{M_{\bullet,2}}

\def\pc{\rm ~pc}
\def\msun{M_{\odot}}
\def\ab{a_{\rm b}}
\def\ah{a_{\rm h}}
\def\abbh{a_{\rm BBH}}
\def\AU{\rm ~AU}
\def\kms{\rm ~km~s^{-1}}
\def\tid{_{\rm tid}}
\def\min{_{\rm min}}
\def\HVS{_{\rm HVS}}
\def\apo{_{\rm apo}}
\def\ini{{\rm ini}}
\def\inf{_{\infty}}

\begin{document}
\shorttitle{Origin of HVSs
}
\shortauthors{ Lu, Zhang, \& Yu}

\title{
On the spatial distribution and the origin of hypervelocity stars
}
\author{Youjun Lu$^1$, Fupeng Zhang$^1$ \& Qingjuan Yu$^{2}$}
\affil{$^1$~National Astronomical Observatories, Chinese Academy of Sciences,
Beijing, 100012, China; luyj,fpzhang@bao.ac.cn \\
$^2$~Kavli Institute for Astronomy and Astrophysics, Peking University, 
     Beijing, 100871, China; yuqj@kiaa.pku.edu.cn}

\begin{abstract}

Hypervelocity stars (HVSs) escaping away from the Galactic halo are dynamical
products of interactions of stars with the massive black hole(s) (MBH) in the
Galactic Center (GC). They are mainly B-type stars with their progenitors
unknown. OB stars are also populated in the GC, with many being hosted in a
clockwise-rotating young stellar (CWS) disk within half a parsec from the MBH
and their formation remaining puzzles. In this paper, we demonstrate that HVSs
can well memorize the injecting directions of their progenitors using both
analytical arguments and numerical simulations, i.e., the ejecting direction of
an HVS is almost anti-parallel to the injecting direction of its progenitor.
Therefore, the spatial distribution of HVSs maps the spatial distribution of
the parent population of their progenitors directly. We also find that almost
all the discovered HVSs are spatially consistent with being located on two thin
disk planes. The orientation of one plane is consistent with that of the
(inner) CWS disk, which suggests that most of the HVSs originate from the CWS
disk or a previously existed disk-like stellar structure with an orientation
similar to it. The rest of HVSs may be correlated with the plane of the
northern arm of the mini-spiral in the GC or the plane defined by the outer
warped part of the CWS disk.  Our results not only support the GC origin of
HVSs but also imply that the central disk (or the disk structure with a similar
orientation) should persist or be frequently rejuvenated over the past 200 Myr,
which adds a new challenge to the stellar disk formation and provides insights
to the longstanding problem of gas fueling into massive black holes.

\end{abstract}

\section{Introduction}\label{sec:intro}

In the past several years, surveys of hypervelocity stars (HVSs) have found 16
HVSs with velocities substantially higher than the Galactic escape velocity
\citep{Brown05,Hirsch05,Edelmann05,Brown07,Brown09a}.  Most HVSs are B-type
stars probably with masses $\sim3-4\msun$, ages $\ga(1-2)\times10^8$~yr, and
lifetime $\sim(1.6-3.5)\times10^8$~yr \citep{Brown07,Brown09a}.  Their
heliocentric distances range from several ten to a hundred kpc and their spatial
distribution on the sky is probably anisotropic
\citep{Brown09a,Brown09b,Abadi09}. The young nature and the spatial anisotropic
distribution of these stars, as well as their hypervelocities, should be
related to their origin.

A few dynamical mechanisms, involving interactions with the massive black hole
(MBH) in the Galactic Center (GC), were proposed to eject stars with such hyper
velocities
\citep{Hills88,YT03,Bromley06,Gualandris05,Levin06,Baumgardt06,Sesana06,Perets07,OL08,LB08},
including tidal breakup of binary star systems by the central MBH and
three-body interactions of single stars with a hypothetical binary black hole
(BBH) in the GC. The sources of (binary) stars injected into the vicinity of
the MBH are often (implicitly) assumed to be isotropically distributed in
previous studies, as the majority of the GC stars are isotropically
distributed; and different ejection mechanisms may result in different spatial
distributions of HVSs, depending on whether the central MBH is a single or a
binary and relevant binary parameters \citep{YT03,Bromley06,Sesana06}. Hence,
the spatial distribution of HVSs was proposed to be useful in identifying
the ejection mechanism responsible for the observed HVSs.\footnote{Some
other observational properties of
HVSs, such as the binarity, rotational velocity and metallicity, were also 
proposed to be useful in identifying the ejection mechanisms of the HVSs
(e.g., \citealt{Luetal07,H07,Przybilla08,LMB08,Perets09a,Perets09b}).}

The progenitors of the discovered B-type young HVSs and their sources, however,
should be distinct populations from the majority of the GC stars (which are
typically old).  Observations have indeed shown various young stellar
structures in the GC, which may be possible parent populations of the HVS
progenitors, including a clockwise young stellar disk (CWS) and the other
possible counterclockwise one within $0.5$~pc from the central MBH
\citep{LB03,Paumard06,LuJ09,Bartko09a}, young stellar clusters like the Arches
and the Quintuplet systems at several ten pc away from the center, and the
tidal remnants of those clusters \citep{Portegies02}.  There also exist other
organized structures in the GC, such as, the circumnuclear molecular disk, the
northern arm (Narm) and the bar components of the minispiral at a few pc from
the center, with which young stars may be associated \citep{YMW00,Paumard06}.
These structures may also be related to the sources of HVSs. All the structures
above are not isotropically distributed, but either planar or orbiting on some
specific planes around the central MBH.  The HVSs, if ejected by interactions
of stars originating from these sources with the central MBH(s), are very
likely to be spatially correlated. And their spatial distribution should be
mapping the distribution of the parent populations of their progenitors, if the
ejected HVSs can well memorize the injecting direction of their progenitors.

This paper is organized as following. In \S~\ref{sec:deflection}, we study how
the spatial distribution of HVSs is related with the geometrical structure of
the parent population of their progenitors. In \S~\ref{sec:Obsconst} we find
that most HVSs discovered so far are probably originated from the clockwise
young stellar disk (CWS) or a disk-like stellar structure with an orientation
similar to that of the CWS disk in the GC. Discussion and conclusions are given
in \S~\ref{sec:disc} and \S~\ref{sec:con}.

\section{The deflection angle: the direction change of HVSs from its progenitor
}\label{sec:deflection}

In this section, we use both analytical analysis and numerical calculations to
demonstrate that HVSs can well memorize the injecting direction of their
progenitors. We do this for different possible dynamical mechanisms of ejecting
HVSs introduced in \S~\ref{sec:intro}.

\subsection{Tidal breakup of a binary star by the central MBH}\label{sec:subBS} 

If a binary star initially unbound or weakly bound to the central MBH
approaches the vicinity of the MBH within a tidal distance
$R\tid\simeq\ab\left(\frac{3\bh}{m\HVS+m_{\rm c}}\right)^{1/3}$, the binary is
probably tidally broken up, and one component of the binary may be ejected as a
HVS, where $\bh$ is the mass of the central MBH [$\simeq4\times10^6\msun$,
\citet{Ghezetal08,Gillessenetal09}], $m\HVS$ and $m_{\rm c}$ are the mass of
the ejected HVS and the other binary component, respectively, and $\ab$ is the
semi-major axis of the binary \citep{Hills88,YT03}.  The initial injecting
velocity of the binary is $\ll\sqrt{G\bh/R\tid}$, where $G$ is the
gravitational constant, and then the velocity of the ejected HVS can be
approximated by \citep{YT03,Bromley06} 
\begin{eqnarray} 
v\inf^{\HVS}&\sim&960\kms\left(\frac{0.6\AU}{{\ab}}\right)^{1/2}\left(\frac{m_{\HVS}+m_{\rm
c}}{8M_{\odot}}\right)^{1/3}\nonumber\\ &&\times\left(\frac{2m_{\rm
c}}{m\HVS+m_{\rm c}}\right)^{1/2}\left(\frac{\bh}{4\times
10^6\msun}\right)^{1/6}.  
\label{eq:vinfbin}\end{eqnarray} 
The orbit of the
ejected HVS and the initial orbit of the injecting binary both have
eccentricities $e$ close to 1 and can be approximated as rectilinear at
distances faraway from the MBH. Below we aim to find out the range of the
deflection angle of these two rectilinear directions.

For convenience, we first consider a purely two-body problem that a star with
mass $m_*$($\ll\bh$) starts at a velocity $v\inf^{\ini}=(G\bh/a)^{1/2}$
from infinity and is moving towards the MBH on a hyperbolic orbit with
periapsis distance $R\min\ll a$. The velocity of the star at periapsis
is $v_{\rm p}=v\inf(2a/R\min+1)^{1/2}$, and then
the deflection angle of its direction from its initial injecting velocity
is $\pi/2-\Psi_1$, where 
$\tan\Psi_1=\sqrt{|1-e^2|}\simeq\sqrt{2R\min/a}\simeq v\inf^{\ini}(2R\min/G\bh)^{1/2}$.
If the star moves towards the MBH on a parabolic or elliptical orbit with
eccentricity close to 1, the deflection angle of the stellar velocity moving
from almost the infinity (excluding locations near the apoapsis) to the
periapsis $R\min$ is about $\pi/2$.  Similarly as above, for a star that can
escape the BH with velocity $v\inf^{\HVS}$, the deflection angle of the
escaping star from its velocity at $R\min$ is $\pi/2-\Psi_2$ with $\tan\Psi_2
\simeq v\inf^{\HVS}(2R\min/G\bh)^{1/2}$.  For the HVSs discovered so far, 
we have $v\inf^{\HVS}\sim750-1000\kms$ obtained by removing the
velocity deceleration due to the Galactic potential measured by
\citet{Xue08}. Note that the relative change of the eccentricity vector
(pointing towards the periapsis from the MBH) is $\sim\delta v_{\rm p}/v_{\rm
p}\sim\Psi_2^2$ and it is negligible compared to $\Psi_2$ for sufficiently
small $\Psi_2$, where $\delta v_{\rm p}\sim(\frac{m_c}{m_{\rm
HVS}+m_c})\sqrt{G(m_{\HVS}+m_c)/\ab}$ is the change of the velocity of
the binary component ejected as the HVS after the binary breakup.  For the
process of tidal breakup of a binary star at a distance of $R_{\rm tid}$ from
the MBH, we approximately have $R\min\sim R_{\rm tid}$, and the total
deflection angle $\Theta$ of the ejected HVS from the original injecting binary
is about $\pi-\sqrt{\Psi_1^2+\Psi_2^2}$. The $\Theta$ is $\sim\pi-(\Psi_1+\Psi_2)$
if the HVS and the injecting binary are on the same
orbital plane and $\sim\pi-\Psi_2$ for $v\inf^{\HVS}\gg v\inf^{\ini}$.
For $\ab\sim0.6\AU$, $v\inf^{\HVS}\sim1000\kms$, and $v\inf^{\rm
ini}\sim250\kms$, we have $\sqrt{\Psi_1^2+\Psi_2^2}\sim0.2\sim10\arcdeg$,
that is, the HVSs are almost reversing the injecting direction of their
progenitors.

\begin{figure}\epsscale{1.00}\plotone{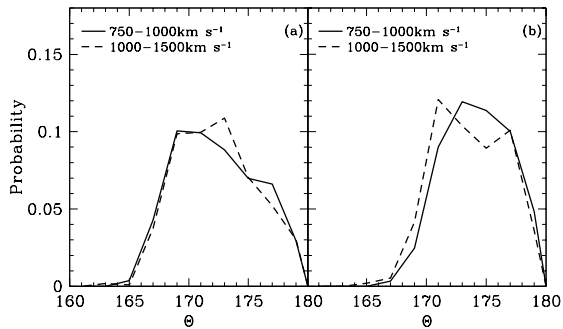}\caption{Probability distribution
of the deflection angle $\Theta$ of HVSs for the mechanism of tidal breakup of
binary stars. Panel a is for the case that the injected binary stars are
initially unbound to the MBH, panel b for the case that the injected binary
stars are initially on weakly bound orbits (see~\S~\ref{sec:subBS}). The solid
and dashed lines represent the distribution of deflection angle for HVSs with
velocities $v_{\infty}^{\HVS}$ in the range of $(750\kms,1000\kms)$ and
$(1000\kms,1500\kms)$, respectively.}\label{fig:f1}\end{figure}

To confirm the above analysis, we numerically realize the process of ejecting an
HVS as a binary interacts with a MBH. We use an explicit 5(4)-order Runge-Kutta
scheme to integrate the full three-body problem \citep{DP80,Haier93}.  In the
three-body simulations, we first assume that all binary stars initially move on
hyperbolic orbits with $v\inf^{\ini}\simeq250\kms$ from infinity.  In
the calculations we set other relevant parameters as follows: (1) The distribution of the
semi-major axes $a_b$ of binary stars is assumed to follow
$P(a_b)da_b\propto\frac{1}{a_b}da_b$ as suggested by observations of binaries
with O-type or B-type primary stars \citep{KF07}. The lower limit of $a_b$ is
roughly set to $0.03\AU$, i.e., about twice the physical radius of a $4\msun$
star; and the upper limit of $a_b$ is set to $2\AU$ to ensure the ejected star
can escape to large Galactic radii.  (2) The mass distribution of primary stars
$m_1$ follows the Miller-Scalo initial mass function, i.e., $f(m_1)\propto
m_1^{-\alpha}$ and $\alpha\sim2.7$ \citep{Kroupa}.  For massive binary stars,
the distribution of the secondary star or the mass ratio $q=m_2/m_1(<1)$ can be
described by two populations: (a) a twin population, i.e., about $40\%$ binary
stars have $q\sim1$ or $m_2\sim m_1$, and (b) the rest binaries follow a
$f(q)\sim$ constant distribution \citep{KF07,Kiminki08,Kiminki09}.  (3) The
eccentricity of the binary star is assumed to be $0$. (4) The
orientation of the binary orbital plane is randomly chosen. (5) The
probability distribution of the closest approach distances to the MBH is
$P(R\min)dR\min\propto dR\min$, which corresponds to the impact parameter
distribution $p(b)db\propto bdb$.  We only select those cases that the masses
of ejected stars are in the mass range $(3\msun,4\msun)$ of the observed HVSs
and calculate the probability distribution of their deflection angles $\Theta$.
As seen from panel a of Figure~\ref{fig:f1},
the values of $\Theta$ range from $165\arcdeg$ to $180\arcdeg$, consistent with
our analysis above.  The distribution of $\Theta$ does not show dependence on
$v_{\infty}^{\HVS}$.  The reason for this independence is as follows. If the
mass of the central MBH (here the MBH in the GC) and the mass of an HVS are
fixed, $v_{\infty}^{\HVS}$ is primarily determined by the semi-major axis of
the binary, and given $R\tid\propto\ab$, $\tan\Psi_2\propto v\inf^{\HVS}(2R\tid/G\bh)^{1/2}$
is independent of $\ab$ and hence $v_{\infty}^{\HVS}$.

We also calculate the deflection angle distribution for the case that the
injected binary stars are initially on weakly bound and highly eccentric
orbits, instead of unbound orbits. In this case, the only difference in the
initial conditions from that for initially unbound binaries is as follows: the
apoapsis distribution of the orbit of the binary barycenter follows
$p(r\apo)dr\apo\propto r\apo^{-1.3}dr\apo$, $r\apo$ is in the range
$(0.04\pc,0.5\pc)$, and the distribution of the closest approach $R\min$, i.e.,
the periapsis distance of the orbit of the binary barycenter, is the same as
above.  The above distribution of the apoapsis is adopted so that the energy
distribution of the binaries injected from a disk structure is consistent with
the density distribution of young stellar disk recently discovered in the GC
\citep{LB03,Paumard06,LuJ09,Bartko09a}.  The radial distribution of stars within
the young stellar disk plane is $\propto r^{-2.3\pm0.7}$ in \citet{LuJ09} and
$\propto r^{-2.1\pm0.2}$ in \citet{Paumard06}.  The resulted distribution of
$\Theta$ is shown in panel b of Figure~\ref{fig:f1}. And $\Theta$ is also in
the range from $165\arcdeg$ to $180\arcdeg$, which again confirms our analysis
above that HVSs memorize the direction of their original binaries well.

\subsection{Dynamical ejection by a binary black hole}\label{sec:subBBH}

HVSs can also be produced by three-body interactions of a single star with
a hard BBH. A BBH is hard if its semi-major axis $\abbh$ is less than 
$\ah=GM_{\bullet,2}/4\sigma^2$, where $\sigma$ is the stellar velocity dispersion
of the host galaxy and $M_{\bullet,2}$ is the mass of the secondary black hole.
For a hard BBH, most low-angular-momentum stars that can enter into the
region $r\la\abbh$ will be ejected after one or several encounters with the BBH
and the r.m.s. of the velocities of the ejected stars at infinity is
\begin{eqnarray} 
v_{\infty}^{\HVS}&\simeq&\sqrt{2KG\bhsp\bhs/(\bh\abbh)}\nonumber\\&\sim&930\kms m_{\bullet}^{0.25}(1-\nu)^{1/2}(0.1\ah/\abbh)^{1/2},\label{eq:1} 
\end{eqnarray}
[see eq.~1 in \citet{Luetal07}], where $M_{\bullet,1}$ is the mass of the
primary black hole, $M_{\bullet}=M_{\bullet,1}+M_{\bullet,2}$, $\nu\equiv\bhs/\bh$,
$m_{\bullet}=\bh/(4\times 10^6\msun)$, $K\simeq1.6$.
Unless the eccentricity of the BBH is excited to an extremely high value shown
in some numerical simulations \citep{Baumgardt06,Makino07,LB08}, the BBH can
stay at its hard stage for a long time (e.g., up to $10^9$~yr in \citealt{YT03})
and ejection of HVSs from the GC would last that long. A BBH with an extremely
high eccentricity has a much shorter lifetime due to gravitation radiation,
which is substantially smaller than the travel time span ($\sim 2\times
10^8$~yr) of the observed HVSs.

\begin{figure}\epsscale{1.00}\plotone{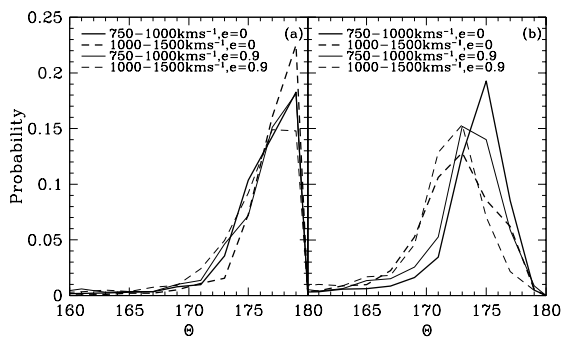}\caption{Probability distribution
of the deflection angle $\Theta$ for the mechanism of three-body interactions
between single stars and a hypothetical BBH in the GC.  The semi-major axis of
the BBH is $0.1\ah$ and its mass ratio is $0.003$.  Panel a is for the case
that the injected single stars are initially unbound to the central BBH, and
panel b for the case that the injected stars are initially on weakly bound
orbits. Different line types have the same meaning as those in Fig.~1.  Thick
lines are for a circular BBH and thin lines for a highly eccentric ($e=0.9$)
BBH.}\label{fig:f2}\end{figure}

In this mechanism, HVSs are ejected out from distances $~R\min\la\abbh$, and
we have $\tan\Psi_2=v_{\infty}^{\HVS}\sqrt{\frac{2R\min}{G\bh}}\la2.5\sqrt{\frac{\bhsp\bhs}{\bh^2}}$.  
If $\bhs\sim\bhsp$, we have $\Psi_2\sim50\arcdeg$, and thus HVSs have lost 
their memory of the direction of their progenitors and the information of the 
BBH orbital plane may be imprinted on the spatial distribution of HVSs.  
Current observational constraints on the mass of the secondary BH in the GC 
gives $\bhs\la0.01\bhsp$ \citep{HM03,YT03,Gillessenetal09}, for which we have
$\Psi_2\la14\arcdeg$.  If the progenitor of the HVS is unbound to the BBH with 
small initial velocities, it is ejected out generally after one or a few close 
encounters with the BBH, and we have
$\tan\Psi_1\la v_{\infty}^{\ini}(\abbh/G\bh)^{1/2}$. For $v_{\infty}^{\rm
ini}\ll v_{\infty}^{\HVS}$, we have $\Psi_1\ll\Psi_2$. If the injecting star is
initially on a bound (and highly eccentric) orbit, it may be ejected out as a
HVS after many times of encounters with the BBH, and the accumulated relative
change of the eccentricity vector is $\sim\Psi_2^2$ and negligible for
sufficiently small $\Psi_2$.  In this case the total deflection angle $\Theta$
of the ejected HVS from the injecting direction of its progenitor is thus
$\pi-\sqrt{\Psi_1^2+\Psi_2^2}\sim\pi-\Psi_2$, and thus HVSs may have a good
memory of the direction of their progenitors.  We do numerical experiments on
three-body interactions between single stars and a BBH with $\bh=4\times10^6\msun$,
$\nu=0.003$, and $\abbh=0.1\ah$, where the two BHs are set to be
sufficiently close so that they are able to eject stars with the observed
hypervelocities. As shown in Figure~\ref{fig:f2}, the deflection angles
$\Theta$ obtained from the calculations range from $165\arcdeg$ to
$180\arcdeg$, no matter whether the injecting stars are initially unbound
(panel a in Fig.~\ref{fig:f2}) or weakly bound (i.e., highly eccentric)
orbits (panel b in Fig.~\ref{fig:f2}). The result is not sensitive to the
BBH eccentricity and $\abbh$.

Hypervelocity stars may also be produced by interactions of a single star with a
stellar-mass black hole in the vicinity of the central MBH. In
this case the velocity kick $\delta v$ received by the star when it passes by
the stellar-mass black hole is typically larger and comparable to the orbital
velocity of the stellar mass black hole $v_{\rm orb}$  and thus the deflection
angle deviates from $\pi$ substantially larger than that for the cases discussed above.  Therefore,
it may be difficult for the HVSs produced by this mechanism to memorize the
direction of their progenitors and hence difficult to explain the consistence
of the HVS plane and the CWS plane shown in \S~\ref{sec:Obsconst} below.

Additional deflection of the moving direction of HVSs may be introduced if the
Galactic potential is substantially flattened or triaxial; but this deflection,
typically $\la2\arcdeg$ for $v_{\infty}^{\HVS}\sim1000\kms$, is negligible
\citep{YM07,Gnedin05}.

In summary, it is plausible to use the spatial distribution of HVSs {\it
unbound} to the Galactic potential to map the parent population of their
progenitors and reveal their origin if the HVSs are produced by the tidal
breakup of binary stars or the BBH mechanism.  

\section{Comparison with observations}\label{sec:Obsconst}

\begin{figure}\epsscale{0.95}\plotone{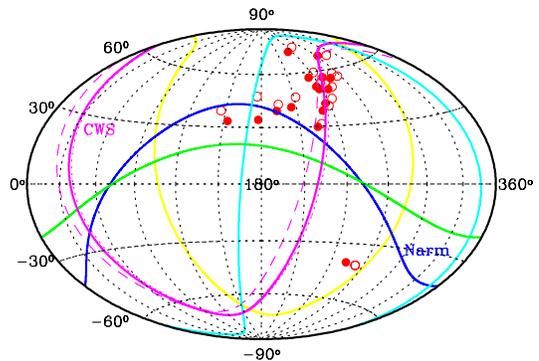}\caption{Hammer-Aitoff projection
in the Galactic coordinates for all the discovered HVSs and some organized
structures in the GC.  Open red circles represent the coordinates (centered on
the sun) of these observed HVSs, while filled red circles represent their
coordinates if projecting the HVS positions to infinity on the sky of an
observer sitting in the GC.  The curves show the planes (also projected to
infinity) of the clockwise young stellar disk (CWS; magenta), the
counterclockwise disk (yellow), the northern arm (Narm; blue) and the bar
(cyan) components of the minispiral, and the circumnuclear disk (green),
respectively \citep{Paumard06}. The magenta curves represent the fitted plane
of the clockwise young stellar disk from \citet{Paumard06} (thick solid curve),
\citet{LuJ09} (thin dashed curve) and \citet{Bartko09a} (thin dotted curve),
respectively.}\label{fig:f3}\end{figure}

Figure~\ref{fig:f3} shows the spatial distribution of all (16) HVSs detected so
far ({\it filled circles}) in the Galactic coordinates by a Hammer-Aitoff
projection.  The spatial distribution of these HVSs is anisotropic at a
3.5-$\sigma$ level \citep{Brown09a,Brown09b} and the majority of the HVSs are
located at Galactic longitudes $l\sim240\arcdeg-270\arcdeg$ \citep{Abadi09}.
The projection of other various planar structures in the GC are also plotted in
the figure.  As shown in Figure~\ref{fig:f3}, most HVSs (11 among 16) situate
close to the plane of the clockwise young stellar disk; and a second group,
including 4 other HVSs and possibly a specific one among the above eleven,
situate close to the Narm plane.  Only one HVS, i.e., HE 0437-5439, is neither
on the CWS plane nor on the Narm plane, which was suggested to be ejected from
the LMC \citep{Edelmann05}.  Almost all HVSs are quite far away from the planes
of other structures plotted in the figure. 

\begin{figure}\epsscale{0.95}\plotone{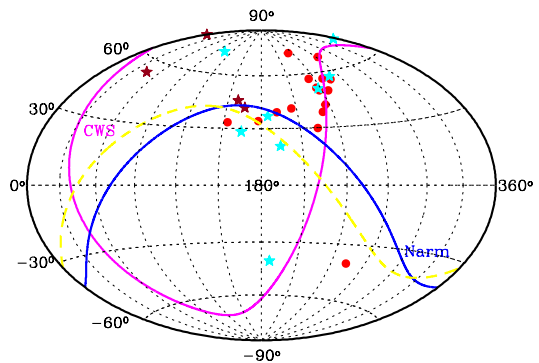}
\caption{
Similar to Fig.~\ref{fig:f3}, with adding the four HVS candidates (carmine
stars), the eight bound HVS (cyan stars), and the plane of the outer warped
part of the CWS disk recently reported by \citet{Bartko09a} (yellow dashed
curve). For clarity, some curves in Fig.~\ref{fig:f3} are removed in this
figure. See details in \S~\ref{sec:Obsconst}.
}
\label{fig:f4}
\end{figure}

We use the $\chi^2$ statistic to fit the possible planar structures of the
HVSs.  The normal $\vec{n}$ to the best-fit plane of the HVSs is obtained by
minimizing $\chi^2\equiv\Sigma_i(\frac{\vec{n}\cdot\vec{e}_{r_i}}{\vec{n}\cdot
d\vec{e}_{r_i}})^2$ ($i=1,2...$), where $\vec{e}_{r_i}$ is the unit position
vector of the $i$-th HVS seen from the GC and $d\vec{e}_{r_i}$ is its possible
deviation from an original plane for which we take equivalently as the error of
$\vec{e}_{r_i}$ in standard statistics.  As the thickness (half-opening angle)
of the CWS plane is $\sim 7\arcdeg-10\arcdeg$ \citep{LuJ09,Bartko09a} and the
deviation angles of ejected HVSs from the direction of their parent population
should be about $\sim 0\arcdeg-15\arcdeg$, we set the amount of
$d\vec{e}_{r_i}$ correspondingly to a deviation of $7\arcdeg$. Thus, we obtain
the normal of the best-fit plane of the eleven HVSs
$(l,b)=(309\arcdeg,-15\arcdeg)$ with $P_{\chi^2}=0.82$, where $P_{\chi^2}$
gives the probability of $\chi^2$ being higher by chance and it is high enough
so that the fit is acceptable.  We take the error of the fitted normal
direction as $\delta l\simeq\pm6\arcdeg$ and $\delta b\simeq\pm8\arcdeg$, which
roughly correspond to the 68\% confidence level.  The normal of the CWS disk
plane, or at least the normal of the CWS disk in the inner region
(0.8\arcsec-3.5\arcsec), $(l,b)=(310\arcdeg,-18\arcdeg)$,
\citep{Paumard06,LuJ09,Bartko09a} are within the error range.  For the
secondary group of five HVSs, we obtain $(l,b)=(188\arcdeg,-52\arcdeg)$ with
$P_{\chi^2}=0.87$, and $(\delta l, \delta b)=(\pm22\arcdeg,\pm5\arcdeg)$
correspond to the 68\% confidence level.  The normal to the observed Narm plane
$(l,b)=(162\arcdeg,-47\arcdeg)$ \citep{Paumard06} is at the 80\% confidence
level of the best-fit value.  

\citet{Bartko09a} recently reported that the outer part of the CWS disk may be
significantly warped and the normal of the warped part is $(l,b)=(136\arcdeg,
-44\arcdeg)$. This normal is close to the normal of the Narm plane as shown in
Figure~\ref{fig:f4}, which suggests some physical connection between the Narm
and the outer warped part of the CWS disk.  The normal of the outer warped part
of the CWS disk is at the 99.9\% confidence level of the best fit. Compared with
the plane of the Narm, statistically the plane defined by the normal of the
warped part of the CWS disk is less likely to be consistent with the fitted
plane of the second population of HVSs.

\begin{figure}\epsscale{1.05}\plotone{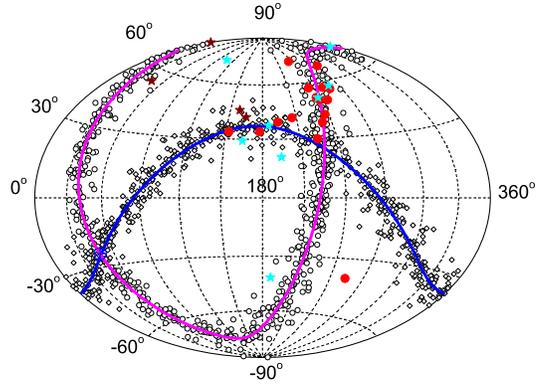}
\caption{
The spatial distribution of simulated HVSs and the spatial distribution of
observed HVSs. Red solid circles, carmine stars and the cyan stars have the
same meanings as those in Fig~\ref{fig:f4}.  Open circles and diamonds
represent the simulated HVSs originated from the two fitted planes with
a thickness of $7\arcdeg$, and normal $(l, b)=(311\arcdeg, -14\arcdeg)$ and
$(176\arcdeg, -53\arcdeg)$, respectively (see details in
\S~\ref{sec:Obsconst}). The magenta and blue curves represent these two best
fitted planes. This figure shows that the spatial distribution of the simulated
HVSs matches that of the observed ones quite well.
}
\label{fig:f5}
\end{figure}

It is worthy to note that there are four other HVS candidates listed in
\citet{Brown09a}. As shown in Figure~\ref{fig:f4}, two of them (SDSS J1403+1450
and SDSS J1546+2437) are within $\sim16\arcdeg$ to the CWS plane and the other
two (SDSS J0940+5309 and SDSS J1014+5631) are within $3\arcdeg$ to the Narm
plane. If they are confirmed to be HVSs and included in the fit, the normals to
the two best-fit planes are $(l,b)=(318\arcdeg,-9\arcdeg)$ with
$P_{\chi^2}=0.81$ and $(l,b)=(180\arcdeg,-50\arcdeg)$ with $P_{\chi^2}=0.79$,
consistent with the planar structures fitted above.  Even for the eight bound
``hypervelocity'' stars listed in \citet{Brown09a}, four of them are close to
the CWS disk within 15$\arcdeg$, so are three of them to the Narm plane (see
Fig.~\ref{fig:f4}). After including them, the normals to the two best-fit planes
are $(l,b)=(311\arcdeg,-14\arcdeg)$ with $P_{\chi^2}=0.44$ and
$(l,b)=(176\arcdeg,-53\arcdeg)$ with $P_{\chi^2}=0.30$, respectively.  As seen
from Figure~\ref{fig:f5}, after including the HVS candidates and the bound
ones, the planar structure close to the CWS disk appears more obvious as their
locations extend to a wider area in the sky.  Only one object among the bound
sample, SDSS J1404+3522, with the smallest Galactocentric distance, is
significantly separate from the above two planes.  We note here that the bound
population of ``hypervelocity'' stars are more likely to be contaminated by the
high-velocity stars produced by other mechanisms as they have substantially
smaller velocities than the unbound stars.  

We have also tested that the fits cannot be passed statistically if choosing
other observed structures, such as, the bar, the circumnuclear molecular disk
or the counter clockwise-rotating disk.  The fit cannot be passed, either, by
fitting all the stars to one best orientation.

Assuming that the parent population of the observed HVSs are on
the two fitted disks with a thickness of $7\arcdeg$ and normal of
$(l, b)=(311\arcdeg, -14\arcdeg)$ and $(176\arcdeg, -53\arcdeg)$, we 
simulate the process of tidal breakup of binary stars around the central
MBH using similar initial conditions as that in \S~\ref{sec:deflection}.
As shown in Figure~\ref{fig:f5}, the spatial distribution of the simulated HVSs
can well match the distribution of observed HVSs. Similar spatial
distribution can also be reproduced if the BBH ejection mechanism is         
alternatively adopted. For simplicity, we do not present that in details. 

\section{Discussion}\label{sec:disc}

As we have demonstrated above, the spatial distribution of the discovered HVSs
is consistent with being located on two thin disk planes, and these two planes
are consistent with that of the  (inner) CWS disk, and the Narm or the
outer warped part of the CWS disk, respectively. We discuss two possible
explanations to these results below. 

One explanation could be that the HVSs are originated from some unknown and
previously existed disk-like stellar structures with orientation similar to that
of the CWS disk, and the Narm or the outer warped part of the CWS disk. If
this is true, one needs to answer what and where the unknown structures are,
why their structures are consistent with the CWS and Narm (or the warped outer
part of the CWS disk), and whether the consistency is coincident or some
natural outcome.

The other explanation is that most HVSs are originated from the CWS disk and a
second population of HVSs may be originated from the Narm or the outer warped
part of the CWS disk.  For this explanation, the young CWS disk should persist
or be frequently rejuvenated over the past $\sim2\times10^8$~yr as constrained
by the travel time of these HVSs [$\sim(1-2)\times10^8$~yr], which is extremely
puzzling.  The ages of the OB stars in the disk are only $\sim6\pm2$~Myr. As
the in-situ formation of the CWS stellar disk is already difficult due to the
suppression of star formation by the MBH tidal field
\citep{LB03,LuJ09,Bartko09a,S92,NC05,BR08}, how have young stars been
continuously forming in the disk and in the meantime how can the disk plane
maintain its direction?  One key to solve this puzzle would be continuous
sinking of cold gas onto the CWS disk. However, the observed gaseous structures
located just outside the CWS disk generally do not have the same direction as
the disk. Solutions to this puzzle will provide profound insights to the gas
fueling into the vicinity of the central MBH in the GC and MBHs in other
galactic nuclei in general. Detailed studies of physical properties of HVSs
(e.g., metallicity) and their relations with those in the CWS stellar disk may
also help to unveil the secret of their star formation. 

It appears that the counter-clockwise disk is not correlated with the currently
discovered HVSs. The current observed B-types in the CWS disk region are also
more isotropic than the O/WR stars \citep{Bartko09b}, which appears not to
be the same as the planar distribution of HVSs.  The correlation between the
observed HVSs and the CWS disk plane may suggest that (1) only the B-type stars
on the disk plane can be perturbed to inject into the immediate vicinity of the
central MBH as the progenitors of HVSs (e.g., by secular evolution of the disk,
see \citealt{MLH09}); and (2) B-type stars initially formed on the disk may be
heated up by relaxation processes later.  Detailed study of this relation may
provide some constraints/hints on the mechanism to deliver stars to the
vicinity of the central MBH. 

It is worthy to further explore how the orbits of (binary) stars in the CWS and
the young stellar structure associated with the Narm plane or the outer warped
part of the CWS disk are perturbed so
that they can move to the immediate vicinity of the central MBH(s). This
perturbation may be due to some massive perturbers \citep{Perets07} or secular
evolution of the structures themselves \citep{MLH09}.  Perturbations on other
young stellar structures, such as the tidal streams of young stellar clusters
like the Arches and the Quintuplet systems, may also inject (binary) stars to
the immediate vicinity of the central MBH(s) and lead to ejection of HVSs.
Therefore, there may be other planar-like spatial distribution of young HVSs.
If these could be found in future HVS surveys, together with those found so
far, the spatial distribution of HVSs will be mapping young stellar structures
ever existed in the GC over the past $2\times10^8$~yr.  With more and more HVSs
to be discovered in the all-sky survey in the future, the statistical methods
on how to extract the stellar structures would need to be improved and the
detailed improvement method should depend on how complicated or simple the
structures would be.

One natural prediction of the HVS origination proposed above is that some
B-type HVSs exist close to the CWS disk plane and the Narm plane or the outer
warped part of the CWS disk in the {\it southern} hemisphere, which should be
a crucial check by future HVS surveys.  

If old-population HVSs can be detected, their spatial distribution may be
different from those shown in Figure~\ref{fig:f3}, as the parent population of
their progenitors may be significantly isotropic than that of the discovered
B-type HVSs. Studying the spatial distribution of different types of HVSs may
help to reveal information on the star formation and the dynamical environment
in the GC.

\section{Conclusions}\label{sec:con}

Using both analytical arguments and numerical simulations, we have demonstrated
that HVSs can well memorize the injecting directions of their progenitors. In
another words, the ejecting direction of an HVS is almost anti-parallel to the
injecting direction of its progenitor. Therefore, the spatial distribution
of HVSs should map the spatial distribution of the parent population of their
progenitors directly. We also find that most of the discovered HVSs are
spatially consistent with being located on two thin disk planes. The
orientation of one plane is consistent with that of the (inner) CWS disk, which
suggests that most of the HVSs originate from it or a disk-like stellar
structure with a similar orientation to it.  The rest of HVSs may be correlated
with the plane of the northern arm of the mini-spiral in the GC or the
plane defined by the outer warped part of the CWS disk. Our results not only
support the GC origin of HVSs but also imply that the central disk (or the disk
structure with a similar orientation) should persist or be frequently
rejuvenated over the past 200 Myr, which adds a new challenge to the stellar
disk formation and provides insights to the longstanding problem of gas fueling
into massive black holes.

\acknowledgements
We thank Scott Tremaine and the referee for helpful comments and Reinhard 
Genzel for discussion on the young stars in the Galactic Center. This work 
was supported in part by the BaiRen program from the National Astronomical 
Observatories, Chinese Academy of Sciences and the National Natural Science 
Foundation of China (10843009). YL and QY thank the hospitality of the Kavli
Institute for Theoretical Physics in the University of California at Santa 
Barbara during their visit, where part of the work was done.

\end{document}